\documentclass{moriond}
\usepackage{xspace}

\def\PRL{\em Phys. Rev. Lett.}
\def\PRD{{\em Phys. Rev.} D}

\def\be{\begin{equation}}
\def\ee{\end{equation}}
\def\bea{\begin{eqnarray}}
\def\eea{\end{eqnarray}}

\def\as{\ensuremath{\alpha_S}\xspace}
\def\ttbar{\ensuremath{t\overline t}\xspace}
\def\mt{\ensuremath{m_t}\xspace}
\usepackage[permil]{overpic}
\usepackage{transparent}

\let\oldsetcounter=\setcounter
\renewcommand\newcommand{\Photo}{\includegraphics[height=35mm]{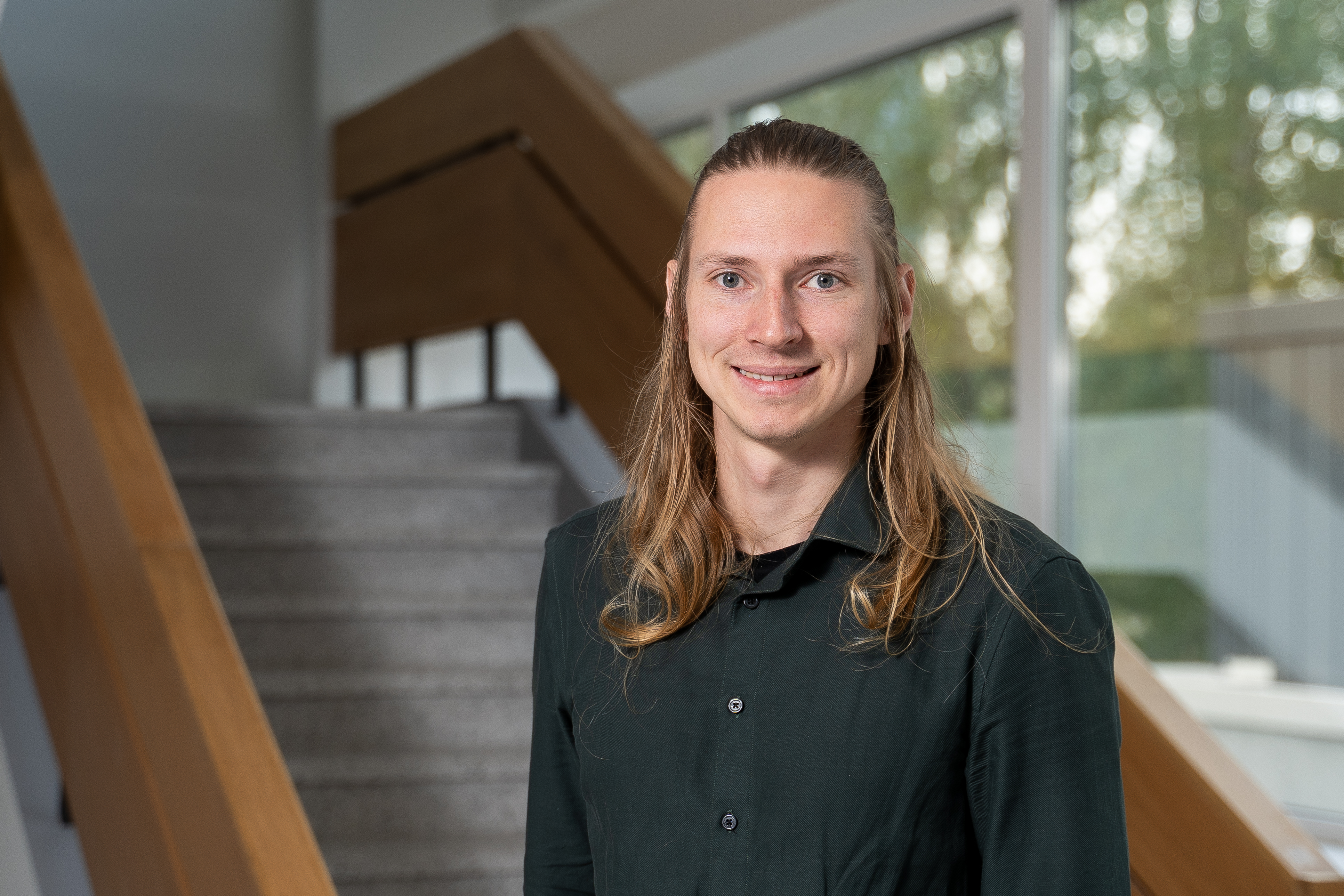}}
\begin{document}
\vspace*{4cm}
\title{Review of the measurements of the
strong coupling constant in CMS at 13 TeV}

\author{ A. Potrebko on behalf of the CMS collaboration }

\address{Institute of Particle Physics and Accelerator Technologies\\
Faculty of Natural Sciences and Technology\\
Riga Technical University, 7 Paula Valdena, Riga LV-1048, Latvia}

\maketitle\abstracts{
The strong coupling constant is the least known of the coupling constants in the standard model. Nevertheless it appears in the calculations of cross sections of all the processes at the LHC. We present a review of the strong coupling constant measurements conducted at the CMS experiment, focusing on those performed at a center-of-mass of 13 TeV.}

\section{Introduction}

\begin{figure}
\begin{minipage}{0.49\linewidth}
\centerline{\includegraphics[width=1.0\linewidth]{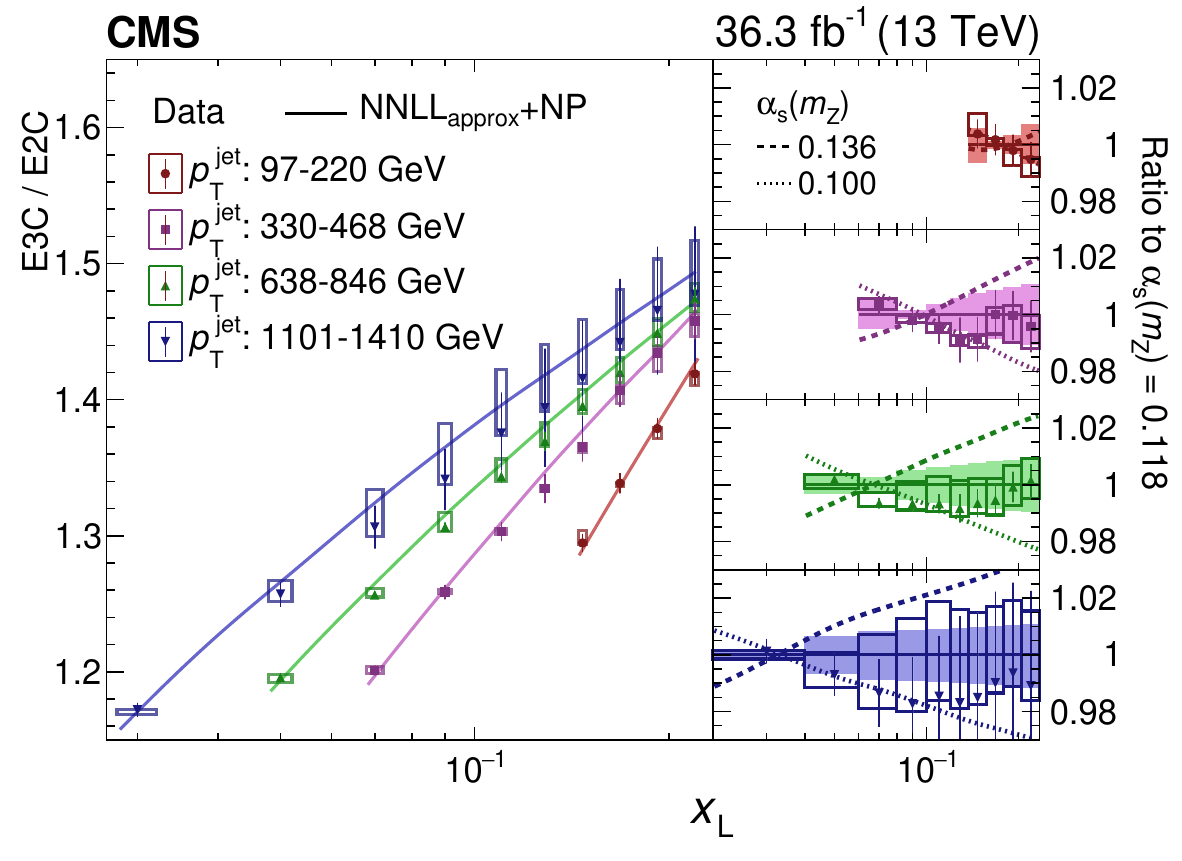}}
\end{minipage}
\hfill
\begin{minipage}{0.49\linewidth}
\centerline{\includegraphics[width=1.13\linewidth]{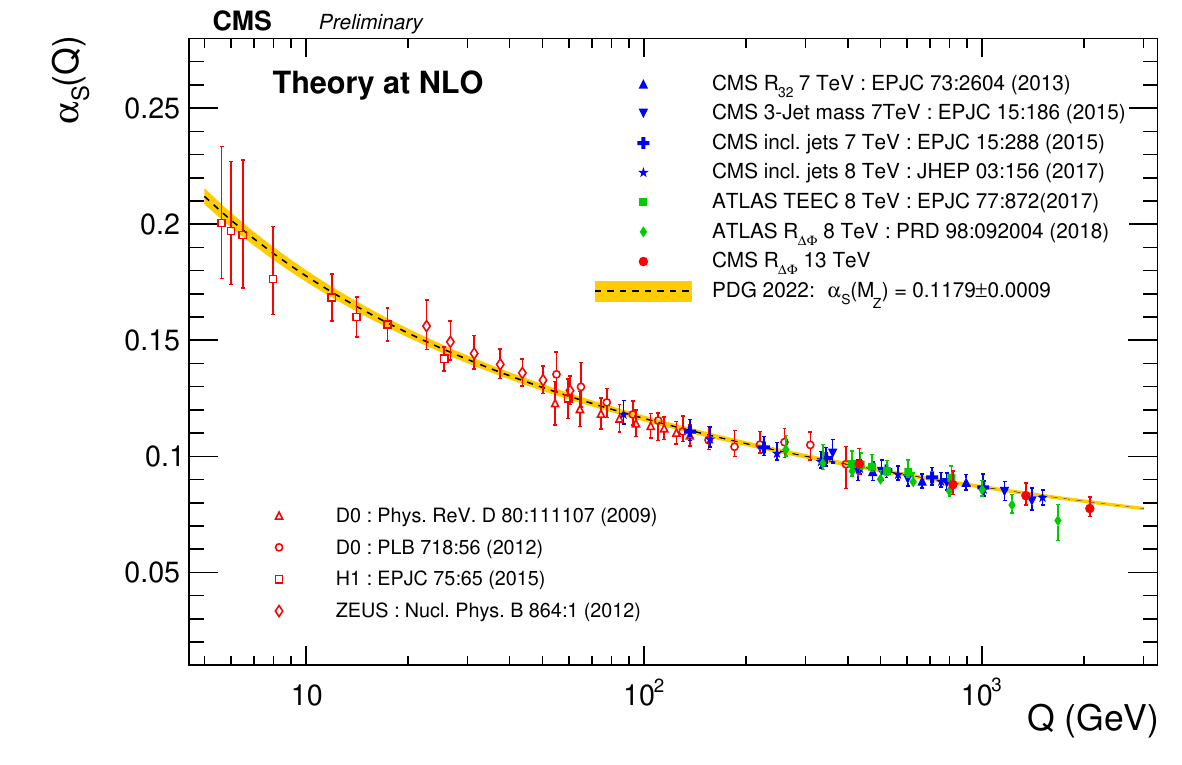}}
\end{minipage}
\hfill
\caption[]{Slope of the ratio of the three- and two-particle energy correlators, E3C/E2C, with respect to the distance between the constituents, $x_L$, for different jet energies \cite{ENC} (left). Unfolded data are overlaid with analytical calculations at $\mathrm{NNLL}_{\mathrm{approx}}$ accuracy. The ratio section illustrates changes in the slope of E3C/E2C for values of \as higher ($\as=0.136$) and lower ($\as=0.100$) than the default value of $\as=0.118$. 
% The ratio observable, $R_{\Delta\phi}$ as a function of jet $p_T$ (right) \cite{Rdphi}. Calculations from pQCD at different \as values are overlaid with unfolded data. REWRITE RIGHT
Running of the strong coupling constant using the world average (yellow band) compared to measurements performed at different scales, $Q$ \cite{Rdphi} (right). Four new extractions from the $R\Delta_{\phi}$ measurement, reaching the highest scale probed to date, are added. 
}
\label{fig:ENC_rdphi}
\end{figure}

The strong coupling constant (\as) is the only free parameter in QCD and is also the least known of the coupling constants in the standard model. The world average value of \as at the reference point of Z boson pole mass is $\as(m_Z)=0.1179\pm0.0009$. Nevertheless, powers of \as appear in the calculations of perturbative quantum chromodynamics (pQCD) for cross sections of virtually all processes measured in LHC. The value of \as decreases with the energy scale of the underlying process, $Q$, as predicted by the renormalization group equations (RGE) starting from the Landau pole value $\Lambda_{QCD}\approx 0.2$ GeV. The consistency of the running of \as with the RGE has been shown to be consistent over three orders of magnitude (see, figure \ref{fig:ENC_rdphi}). In the CMS experiment \cite{CMS_exp} \as has been measured in vector boson, top, jet and jet substructure measurements. In this review, we summarize the measurements performed in the CMS at the center-of-mass energy $\sqrt{s}=13$ TeV, first highlighting three new measurements.

\section{New measurements of the strong coupling constant in CMS} \label{label:new}
A novel approach in HEP is to measure \as using energy correlators inside jets \cite{ENC}. N-particle energy correlators (ENC) describe the correlations of kinematic properties of particles inside jets. In the new measurement, two- and three-particle energy correlators (E2C and E3C, respectively), were measured. E2C (E3C) sum up all the combinations of pairs (triplets) of particles inside jets, scaled with an energy weight. For example,
\begin{equation}\label{eq:E3C}
    \mathrm{E3C} = \sum_{i,j,k} \int d\sigma \frac{E_i E_j E_k}{E^3} \delta \Bigl(x_L-\mathrm{max}(\Delta R_{i,j}, \Delta R_{i,k}, \Delta R_{j,k})\Bigr),
\end{equation}
where $\Delta R_{i,j} = \sqrt{(\Delta \eta_{i,j})^2 + (\Delta \phi_{i,j})^2}$ represents the angular distance between two constituents. The dependence of E2C and E3C on $x_L$ reveals mappings of various stages of parton showers. Short $x_L$ describe the final stages of fragmentation, that is free hadrons that are uncorrelated. The scaling is linear in this case. Conversely, large $x_L$ describe interacting partons, characterized by an inverse scaling.

In this measurement, dijet events with jet rapidity $|\eta^{jet}|<2.1$ and transverse momentum $p_T^{jet}>97$ GeV were selected. In jets, all charged and neutral particles with $p_T>1$ GeV were selected. The data used for the extraction of \as were corrected for the detector effects (unfolded). The ratio of energy correlators $\mathrm{E2C/E3C} ~\sim \alpha \mathrm{log} R$. In this way for the extraction of \as, $\mathrm{E2C/E3C}$ was compared to analytical calculations available at an approximate next-to-next-to-leading logarithm ($\mathrm{NNLL}_{\mathrm{approx}}$) matched to next-to-leading order (NLO) perturbative QCD (pQCD) calculations. This yielded the worlds most precise \as measurement from jet substructure: $\alpha_S=0.1229^{+0.0040}_{-0.0050}$. While the uncertainty of the measurement is larger compared to measurements obtained from jet and top quark cross section measurements, \as obtained from jet substructure is more sensitive to collinear effect. In this way, this measurement helps to probe the consistency of \as in different phase spaces. 

Another new measurement of \as at CMS was performed using azimuthal correlations among jets \cite{Rdphi,Rdph2}.  A ratio observable 
\begin{equation} \label{eq:DeltaRphi}
    R_{\Delta \phi} (p_T) = \frac{\sum_{i=0}^{N_{jet}(p_T)} N_{nbr}^{(i)} (\Delta\phi, p_{T,min}^{nbr}) }{N_{jet} (p_T)}
\end{equation} was defined, where the denominator counts the number of jets pet $p_T$ bin while the numerator counts the number of neighboring jets for a given jet $i$. Neighboring jets in this measurement were taken to fall within the interval $\frac{2\pi}{3} < \Delta \phi < \frac{7\pi}{8}$, i.e., neighboring jets do not have to be spatially close to each other. Such a choice of $\Delta \phi$ ensures that in the dijet event, each jet has 0 neighbors, resulting in the numerator of $R_{\Delta\phi}$ counting only the 3+ jet topologies while the denominator counts all the jets. 

% \begin{figure}
% % \begin{minipage}{0.45\linewidth}
% \centerline{\includegraphics[width=0.5\linewidth]{pics/Rdphi_running.pdf}}
% % \end{minipage}
% \caption[]{}
% \label{fig:radish}
% \end{figure}

For the extraction of \as, perturbative QCD calculation using NLOJet++ within fastNLO framework was compared to the unfolded data. Non-perturbative (NP) effects for data were estimated using Monte Carlo (MC) predictions taking a ratio of the distributions with and without NP effects $C^{NP} = (\sigma^{PS+MPI+HAD})/\sigma^{PS}$. NLO electroweak (EW) effects were estimated from the SHERPA MC generator in interfaced to RECOLA. The fit yielded $\as(m_Z) = 0.1177^{+0.0117}_{-0.0074}$. The main contribution to the uncertainty is the scale uncertainty ($^{+114}_{-68}$) which could be reduced by around threefold if next-to-NLO (NNLO) predictions were available. Experimental, NP and EW uncertainties are small due to cancellations when taking the ratio. The measurement also probed running of \as, splitting the whole $p_T$ range into four subregions. Notably, the measurement reaches to the highest scale to date, $Q\approx 2081$ TeV showing no deviations from RGE.   

In the multi-differential dijet cross section measurement, a double-differential (2D) cross section as a function of the invariant mass of the two jets, $m_{1,2}$, and the largest absolute rapidity, $|y|_{max}$, was used to extract \as\cite{dijet}. A simultaneous fit of parton distribution functions (PDF) and \as was performed by fitting the unfolded data to pQCD predictions available up to NNLO. The predictions were obtained with the NNLOJET program interfaced with fastNLO. The fit resulted in $\as(m_Z)=0.1179\pm0.0019$. A triple-differential (3D) distribution as a function of the rapidity separation, $y^*$, the total boost, $y_b$, and $m_{1,2}$ was also used to extract \as and yielded $\as=0.1181\pm0.022$ which is in a good agreement with the 2D result. The main uncertainty in this measurement is the fit, i.e., the experimental uncertainty mostly caused by the jet energy scale uncertainty and luminosity.

\section{Summary of the measurements of the strong coupling constant in CMS}

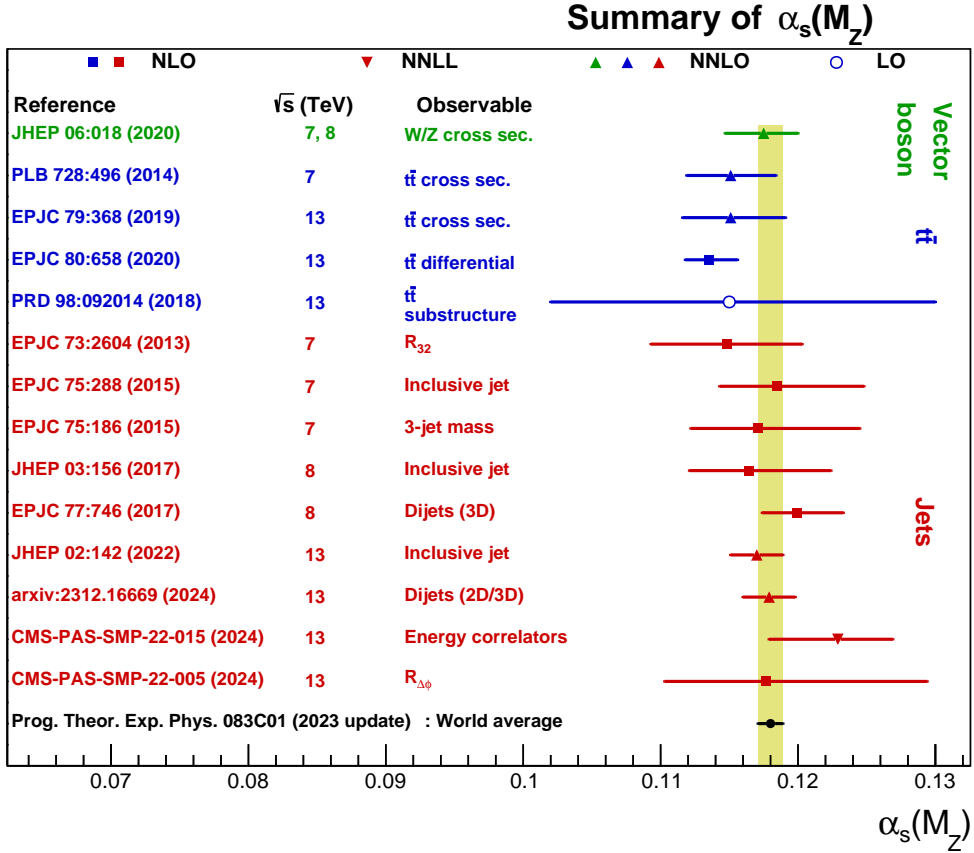
\begin{figure}
\begin{overpic}[width=0.99\linewidth]{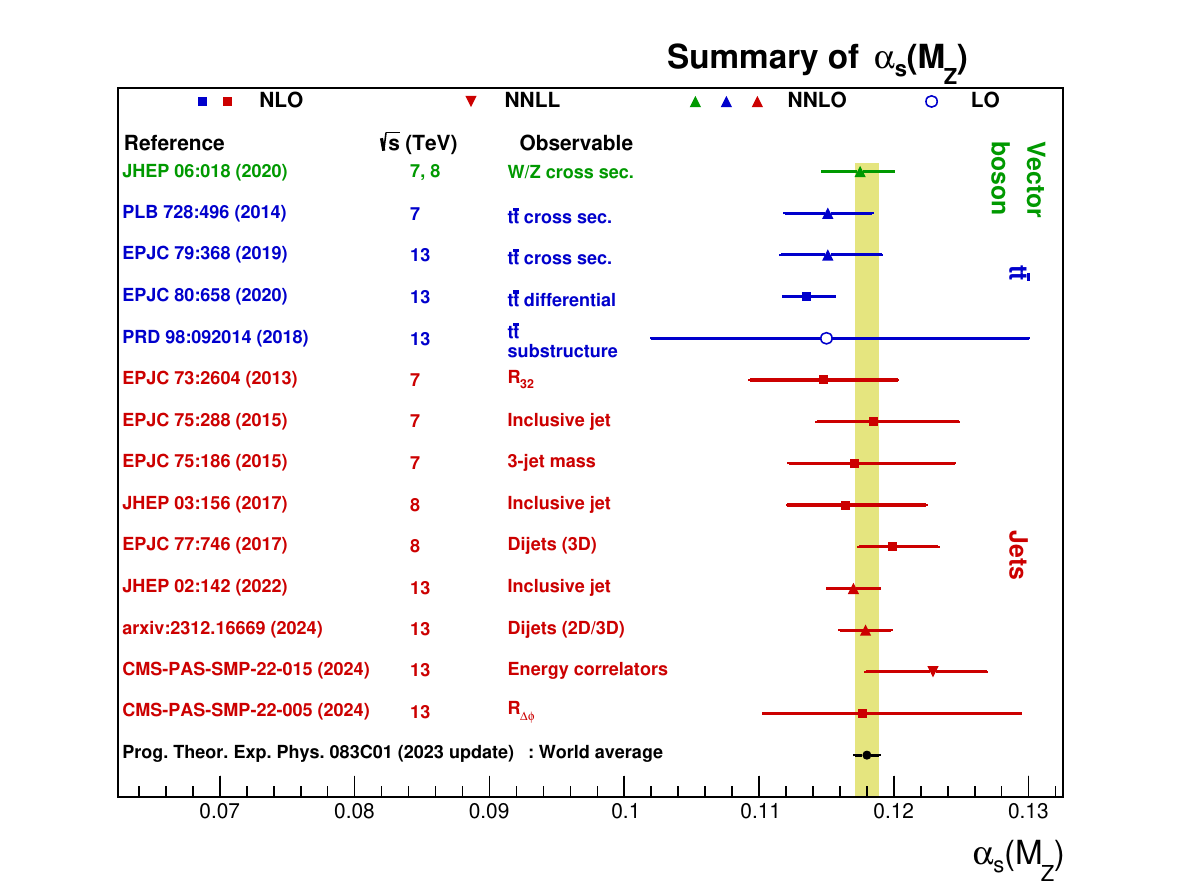}
% \href{}{\put(450,450){\linethickness{0.25mm}\color{black}\polygon(0,0)(290,0)(290,120)(0,120)}}%
% \put(100, 600){\transparent{0.2}\href{https://cms-results.web.cern.ch/cms-results/public-results/publications/TOP/index.html}{\colorbox{white}{Just some text}}}
\put(90,585){
  \href{https://link-springer-com.ezproxy.cern.ch/article/10.1007/JHEP06(2020)018}{
    \color{red}\transparent{0}\rule{34mm}{16px}}
}
\put(90,559){
  \href{https://www.sciencedirect.com/science/article/pii/S0370269313009854}{
    \color{red}\transparent{0}\rule{34mm}{14px}}
}
\put(90,520){
  \href{https://link-springer-com.ezproxy.cern.ch/article/10.1140/epjc/s10052-019-6863-8}{
    \color{red}\transparent{0}\rule{34mm}{15px}}
}
\put(90,485){
  \href{https://link-springer-com.ezproxy.cern.ch/article/10.1140/epjc/s10052-020-7917-7}{
    \color{red}\transparent{0}\rule{34mm}{15px}}
}
\put(90,450){
  \href{https://journals-aps-org.ezproxy.cern.ch/prd/abstract/10.1103/PhysRevD.98.092014}{
    \color{red}\transparent{0}\rule{34mm}{14px}}
}
% \put(90,450){
%   \href{https://cms-results.web.cern.ch/cms-results/public-results/publications/TOP/index.html}{
%     \color{red}\transparent{0}\rule{34mm}{22mm}}
% }
\put(90,130){
  \href{https://cms-results.web.cern.ch/cms-results/public-results/publications/SMP/ALPHAS.html}{
    \color{red}\transparent{0}\rule{34mm}{49mm}}
}
\end{overpic}
\caption[]{Summary of the \as$(m_Z)$ measurements in CMS and the world average (black maker and the yellow band). The error bands illustrate the full uncertainty of the measurement.}
\label{fig:alpha_s}
\end{figure}

A summary of the measurements of \as in CMS at $\sqrt{s} = $ 7, 8 and 13 TeV can be seen in figure \ref{fig:alpha_s}. In this section, we will review the measurements of \as at $\sqrt{s}=13$ Tev, not mentioned in section \ref{label:new}. 

In the inclusive jet measurement \cite{incl_jet}, \as was extracted from a fit to a 2D cross section of jet $p_T$ and $|\eta|$. A simultaneous fit of PDF and \as was performed using NNLO predictions corrected for NP and EW effects. The value obtained is $\alpha_s(m_Z) = 0.1166\pm0.0017$. The uncertainty is dominated by the fit uncertainty primarily stemming from the jet energy scale uncertainty.

When fitting \as using top quark datasets, the simultaneous fit should also include top mass, $m_t$. However, in the inclusive cross section measurement only one parameter can be used to extract $\alpha_s$. In the top quark pair (\ttbar) inclusive cross section measurement in the two lepton decay channel, instead of a simultaneous fit, $\alpha_s$ was extracted several times for different PDF sets using $m_t$ default to each PDF set \cite{inclusive_ttbar}. The obtained value, $\alpha_s=0.1151^{+0.0040}_{0.0035}$, is consistent with the world average. In addition, the fit stability was assessed by repeating the fit for different $m_t$ values. The result for \as was found to deviate by no more than one sigma if $m_t$ default to other PDF sets was used instead.  

In the multi-differential \ttbar cross section measurement, \as was extracted from a 3D distribution of the number of jets in the event, $N_{jet}$, mass and the absolute rapidity of the \ttbar system, $M(\ttbar)$ and  $\eta(\ttbar)|$ respectively \cite{multidiff_ttbar}. A simultaneous fit of \as, $m_t$ and PDF was performed compared to NLO calculations. The value of $\as(m_Z) = 0.1135^{+0.0021}_{-0.0017}$ and $\mt=170.5\pm0.8$ GeV was obtained. In the measurement possible effects from Coulomb and soft-gluon resummation at the \ttbar threshold were neglected. This would cause an increase of \mt by an order of 1 GeV and would also pull \as by an order of 0.001 to a higher value. Despite the absence of this contribution, the competitive uncertainty suggests that the data should be reexamined. While the given measurement only analyzed the CMS data acquired in 2016, a new measurement extended to the full Run 2 data is now available and can be included in the extraction of \as \cite{multidiff_ttbar_full}.

Finally \as was extracted from bottom jet substructure in \ttbar events \cite{jet_substr}. In this measurement, the angle between the groomed subjects, $\Delta R_g$, was used for \as extraction. The data were fitted with the predictions generated by the POWHEG generator and showered using the PYTHIA~8 program. This only provides LO+LL precision for distributions within jets, resulting in a large uncertainty in the final measurement $\as = 0.115^{+0.015}_{-0.013}$. Soft-gluon emissions in this measurement were incorporated using the CMW scheme.

\section{Summary}
At $\sqrt{s}=13$ TeV, the CMS Collaboration has provided 7 different \as extractions spanning jet, top and jet substructure measurements. Where NNLO calculation are available the experimental fit uncertainties are typically the dominant ones. An option to mitigate this is to use ratio observables like $R_{\Delta \phi}$ or $R_{32}$ where several uncertainties cancel out. Conversely, measurements like $R_{\Delta \phi}$ and jet substructure would greatly benefit from higher-order predictions.

\section*{References}

% \bibliography{MoriondQCD24}

\end{document}